# A New Approach to Keyphrase Extraction Using Neural Networks

**Kamal Sarkar, Mita Nasipuri and Suranjan Ghose**

**Computer Science and Engineering Department, Jadavpur University,**
**Kolkata-700 032, India**

**Abstract**
Keyphrases provide a simple way of describing a document, giving the reader some clues about its contents. Keyphrases can be useful in a various applications such as retrieval engines, browsing interfaces, thesaurus construction, text mining etc.. There are also other tasks for which keyphrases are useful, as we discuss in this paper. This paper describes a neural network based approach to keyphrase extraction from scientific articles. Our results show that the proposed method performs better than some state-of-the art keyphrase extraction approaches.

## 1. Introduction

The pervasion of huge amount of information through the World Wide Web (WWW) has created a growing need for the development of techniques for discovering, accessing, and sharing knowledge. The keyphrases help readers rapidly understand, organize, access, and share information of a document. Keyphrases are the phrases consisting of one or more significant words. keyphrases can be incorporated in the search results as subject metadata to facilitate information search on the web [1]. A list of keyphrases associated with a document may serve as indicative summary or document metadata, which helps readers in searching relevant information.

Keyphrases are meant to serve various goals. For example, (1) when they are printed on the first page of a journal document, the goal is summarization. They enable the reader to quickly determine whether the given article worth in-depth reading. (2) When they are added to the cumulative index for a journal, the goal is indexing. They enable the reader to quickly find a article relevant to a specific need. (3) When a search engine form contains a field labeled keywords, the goal is to enable the reader to make the search more precise. A search for documents that match a given query term in the keyword field will yield a smaller, higher quality list of hits than a search for the same term in the full text of the documents. When the searching is done on the limited display area devices such as mobile, PDA etc. , the concise summary in the form of keyphrases , provides a new way for displaying search results in the smaller display area[ 2] [ 3].

Although the research articles published in the journals generally come with several author assigned keyphrases, many documents such as the news articles, review articles etc. may not have author assigned keyphrases at all or the number of author-assigned keyphrases available with the documents is also too limited to represent the topical content of the articles. Many documents also do not come with author assigned keyphrases. So, an automatic keyphrase extraction process is highly desirable.

Manual selection of keyphrases from a document by a human is not a random act. Keyphrase extraction is a task related to the human cognition. Hence, automatic keyphrase extraction is not a trivial task and it needs to automated due to its usability in managing information overload on the web.

Some previous works on automatic keyphrase extraction used the machine learning techniques such as Naïve Bayes, Decision tree, genetic algorithm [15] [16] etc.

Wang et.al (2006) has proposed in [14] a neural network based approach to keyphrase extraction, where keyphrase extraction has been viewed as a crisp binary classification task. They train a neural network to classify whether a phrase is keyphrase or not. This model is not suitable when the number of phrases classified by the classifier as positive is less than the desired number of keyphrases, K.

To overcome this problem, we think that keyphrase extraction is a ranking problem rather than a classification problem. One good solution to this problem is to train a neural network to rank the candidate phrases. Designing such a neural network requires the keyphrases in the training data to be ranked manually. Sometimes, this is not feasible.

In this paper, we present a keyphrase extraction method that uses a multilayer perceptron neural network which is trained to output the probability estimate of a class: positive (keyphrase) or negative (not a keyphrase). Candidate phrases which are classified as positive are ranked first based on their class probabilities. If the number of desired keyphrases is greater than the number of phrases classified as positive by the classifier, the candidate phrases classified as negative by the classifier are considered and they are sorted in increasing order of the their class probabilities, that is, the candidate phrase classified as negative with minimum probability estimate is added first to the list of previously selected Keyphrases. This process continues until the number of extracted keyphrases exceed the number K, where K = the desired number of the keyphrases.

Our work also differs from the work proposed by Wang et.al (2006) [14] in the number and the types of features used. While they use the traditional TF*IDF and position features to identify the keyphrases, we use extra three features such as phrase length, word length in a phrase, links of a phrase to other phrases. We also use the position of a phrase in a document as a continuous feature rather than a binary feature.

The paper is organized as follows. In section 2 we present the related work. Some background knowledge about artificial neural network has been discussed in section 3. In section 4, the proposed keyphrase extraction method has been discussed. We present the evaluation and the experimental results in section 5.

## 2. Related Work

A number of previous works has suggested that document keyphrases can be useful in a various applications such as retrieval engines [1], [4], browsing interfaces [5], thesaurus construction [6], and document classification and clustering [7].

Some supervised and unsupervised keyphrase extraction methods have already been reported by the researchers. An algorithm to choose noun phrases from a document as keyphrases has been proposed in [8]. Phrase length, its frequency and the frequency of its head noun are the features used in this work. Noun phrases are extracted from a text using a base noun phrase skimmer and an off-the-shelf online dictionary.

Chien [9] developed a PAT-tree-based keyphrases extraction system for Chinese and other oriental languages.

HaCohen-Kerner et al [10][11] proposed a model for keyphrase extraction based on supervised machine learning and combinations of the baseline methods. They applied J48, an improved variant of C4.5 decision tree for feature combination.

Hulth et al [12] proposed a keyphrase extraction algorithm in which a hierarchically organized thesaurus and the frequency analysis were integrated. The inductive logic programming has been used to combine evidences from frequency analysis and thesaurus.

A graph based model for keyphrase extraction has been presented in [13]. A document is represented as a graph in which the nodes represent terms, and the edges represent the co-occurrence of terms. Whether a term is a keyword is determined by measuring its contribution to the graph.

A Neural Network based approach to keyphrase extraction has been presented in [14] that exploits traditional term frequency, inverted document frequency and position (binary) features. The neural network has been trained to classify a candidate phrase as keyphrase or not.

Turney [15] treats the problem of keyphrase extraction as supervised learning task. In this task, nine features are used to score a candidate phrase; some of the features are positional information of the phrase in the document and whether or not the phrase is a proper noun. Keyphrases are extracted from candidate phrases based on examination of their features. Turney's program is called Extractor. One form of this extractor is called GenEx, which is designed based on a set of parameterized heuristic rules that are fine-tuned using a genetic algorithm. Turney Compares GenEX to a standard machine learning technique called Bagging which uses a bag of decision trees for keyphrase extraction and shows that GenEX performs better than the bagging procedure.

A keyphrase extraction program called Kea, developed by Frank et al. [16][17], uses the Bayesian learning technique for keyphrase extraction task. A model is learned from the training documents with exemplar keyphrases and corresponds to a specific corpus containing the training documents. Each model consists of a Naive Bayes classifier and two supporting files containing phrase frequencies and stopped words. The learned model is used to identify the keyphrases from a document. In both Kea and Extractor, the candidate keyphrases are identified by splitting up the input text according to phrase boundaries (numbers, punctuation marks, dashes, and brackets etc.). Finally a phrase is defined as a sequence of one, two, or three words that appear consecutively in a text. The phrases beginning or ending with a stopped word are not taken under consideration. Kea and Extractor both used supervised machine learning based approaches. Two important features such as distance of the phrase's first appearance into the document and TF*IDF (used in information retrieval setting), are considered during the development of Kea. Here TF corresponds to the frequency of a phrase into a document and IDF is estimated by counting the number of documents in the training corpus that contain a phrase P. Frank et al. [16][17], has shown that the performance of Kea is comparable to GenEx proposed by Turney.

An n-gram based technique for filtering keyphrases has been presented in [18]. In this approach, authors compute n-grams such as unigram, bigram etc for extracting the candidate keyphrases which are finally ranked based on the features such as term frequency, position of a phrase in a document and a sentence.

## 3. Background

In this section, we briefly describe some basics of artificial neural network and how to estimate class probability in an artificial neural network. The estimation of class probabilities is important for our work because we use the estimated class probabilities as the confidence scores which are used in re-ranking the phrases belonging to a class: positive or negative.

Artificial Neural networks (ANN) are predictive models loosely motivated by the biological neural systems. In generic sense, the terms "Neural Network" (NN) and "Artificial Neural Network" (ANN) usually refer to a Multilayer Perceptron (MLP) Network, which is the most widely used types of neural networks. A multiplayer perceptron (MLP) is capable of expressing a rich variety of nonlinear decision surfaces. An example of such a network is shown in Figure 1. A multilayer perceptron neural network has usually three layers: one input layer, one hidden layer and one output layer. A vector of predictor variable values ($x_1...x_i$) is presented to the input layer. In the keyphrase extraction task, this input vector is the feature vector, which is a vector of values of features characterizing the candidate phrases. Before presenting a vector to the input layer, it is normalized. The input layer distributes the values to each of the neurons in the hidden layer. In addition to the predictor variables, there is a constant input of 1.0, called the bias that is fed to each of the hidden layers. The bias is multiplied by a weight and added to the sum going into the neuron. The value from each input neuron is multiplied by a weight ($w_{ij}$) and arrives at a neuron in the hidden layer, and the resulting weighted values are added together producing a combined value at a hidden node. The weighted sum is then fed into a transfer function (usually a sigmoid function), which outputs a value. The outputs from the hidden layer are distributed to the output layer. Arriving at a node (a neuron) in the output layer, the value from each hidden layer neuron is again multiplied by a weight ($w_{jk}$), and the resulting weighted values are added together producing a combined value at an output node. The weighted sum is fed into a transfer function (usually a sigmoid function), which outputs a value $O_k$. The $O_k$ values are the outputs of the network.

One hidden layer is sufficient for nearly all problems. In some special situations such as modeling data which contains a saw tooth wave like discontinuities, two hidden layers may be required. There is no theoretical reason for using more than two hidden layers.

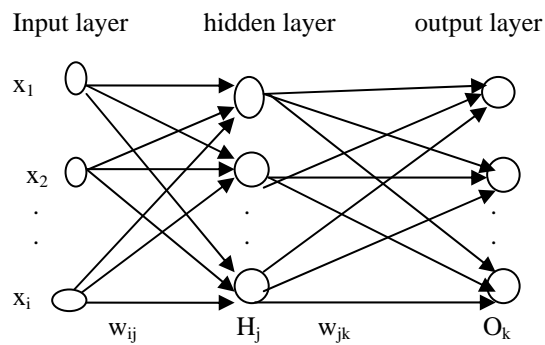

Fig.1. A multilayer feed-forward neural network: A training sample, $X = (x_1, x_2, . . . x_i)$, is fed to the input layer. Weighted connections exist between each layer, where $w_{ij}$ denotes the weight from a unit $j$ in one layer to a unit $i$ in the previous layer.

The backpropagation algorithm performs learning on a multilayer feed-forward neural network. The backpropagation training algorithm was the first practical method for training multiplayer perceptron (MLP) neural networks. The backpropagation (BP) algorithm implements a gradient descent search through the space of possible network weights, iteratively reducing the error between the training example target values and network outputs. BP allows supervised mapping of input vectors and corresponding target vectors. The backpropagation training algorithm follows the following cycle to refine the weight values:

(1) randomly choose a tentative set of weights (initial weight configuration) and run a set of predictor variable values through the network, (2) compute the difference between the predicted target value and the training example target value, (3) average the error information over the entire set of training instances, (4) propagate the error backward through the network and compute the gradient (vector of derivatives) of the change in error with respect to changes in weight values, (5) make adjustments to the weights to reduce the error. Each cycle is called an epoch.

One of the most important issues in designing a perceptron network is the number of neurons to be used in the hidden layer(s). If an inadequate number of neurons are used, the network will be unable to model complex data, and the resulting network will fit poorly to the training data. If too many neurons are used, the training time may be excessively long, and the network may over fit the data. When overfitting occurs, the network will begin to model random noise in the data. As a result, the model fits the training data extremely well, but it performs poorly to new, unseen data. Cross validation can be used to test for this. The number of neurons in the hidden layers may be optimized by building models using varying numbers of neurons and measuring the quality using cross validation method.

### 3.1 Computing Class probability

Given the training data, the standard statistical technique such as Parzen Windows [22] can used to estimate the probability density in the output space. After calculating the output vector O for an unknown input, one can compute the estimated probability that it belongs to each class using the following formula:

$$P_{co}(c|O) = \frac{p(c|O)}{\sum_{c'} p(c'|O)}, \text{ for class c}$$

p(c|O) is the density of points of the category C at location O in the scatter plot of category 1 Vs. Category 0 in a two class problems [23].

We use the estimated class probabilities as the confidence scores to order phrases belonging to a class: positive or negative.

## 4. Proposed Keyphrase Extraction Method

The proposed keyphrase extraction method consists of three primary components: document preprocessing, candidate phrase identification and keyphrase extraction using a neural network.

### 4.1 Document Preprocessing

The preprocessing task includes formatting each document. If a source document is in pdf format, it is converted to a text format before submission to the keyphrase extractor.

### 4.2 Candidate Phrase Identification

The candidate phrase identification is an important step in key phrase extraction task. We treat all the noun phrases in a document as the candidate phrases [1]. The following sub-section discusses how to identify noun phrases.

**Noun Phrase Identification**

To identify the noun phrases, documents should be tagged. The articles are passed to a POS tagger called MontyTagger [25] to extract the lexical information about the terms. Figure 2 shows a sample output of the Monty tagger for the following text segment:

"European nations will either be the sites of religious conflict and violence that sets Muslim minorities against secular states and Muslim communities against Christian neighbors, or it could become the birthplace of a liberalized and modernized Islam that could in turn transform the religion worldwide."

European/JJ nations/NNS will/MD either/DT be/VB the/DT sites/NNS of/IN religious/JJ conflict/NN and/CC violence/NN that/IN sets/NNS Muslim/NNP minorities/NNS against/IN secular/JJ states/NNS and/CC Muslim/NNP communities/NNS against/IN Christian/NNP neighbors/NNS,/, or/CC it/PRP could/MD become/VB the/DT birthplace/NN of/IN a/DT liberalized/VBN and/CC modernized/VBN Islam/NNP that/WDT could/MD in/IN turn/NN transform/VB the/DT religion/NN worldwide/JJ ./.

Fig.2 A sample output of the tagger

In figure 2, NN,NNS,NNP,JJ,DT,VB,IN,PRP,WDT,MD etc. are lexical tags assigned by the tagger.

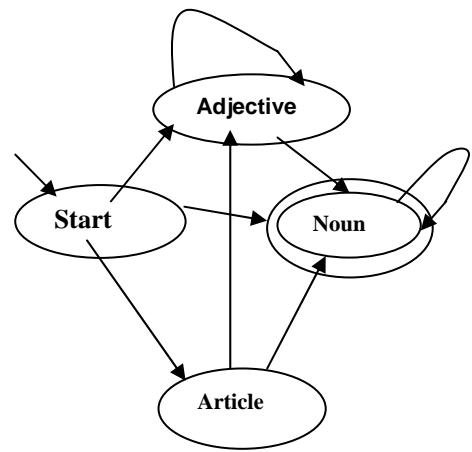

Fig. 3 DFA for noun phrase identification

The meanings of the tags are as follows:

NN and NNS for nouns (singular and plural respectively), NNP for proper nouns, JJ for adjectives, DT for determiner, VB for a verb, IN for a preposition, PRP for a pronoun. This is not the complete tag set.

The above mentioned tags are some examples of tags in the Penn Treebank tag set used by the MontyTagger.

The noun phrases are identified from the tagged sentences using the DFA (deterministic finite automata) shown in figure 3. In this DFA, the states for adjective, noun represent all variations of adjectives and nouns.

The figure 4 shows the noun phrases identified by our noun phrase identification component when the tagged sentences shown in figure 2 become its input. As shown in the figure 4, the 10$^{th}$ phrase is "Islam", but manual inspection of the source text may suggest that it should be "Modernized Islam". This discrepancy occurs since the tagger assigns a tag "VBN" to the word "Modernized" and "VBN" indicates participle form of a verb which is not accepted by our DFA in figure 3 as the part of a noun phrase. To avoid this problem "VBN" might be considered as a state in the DFA, but it might lead to recognizing some verb phrases mistakenly as the noun phrases.

| Document number | Sentence Number | Noun phrase Number | Noun Phrases |
|---|---|---|---|
| 100 | 4 | 1 | European nations |
| 100 | 4 | 2 | sites |
| 100 | 4 | 3 | religious conflict |
| 100 | 4 | 4 | violence |
| 100 | 4 | 5 | sets muslim minorities |
| 100 | 4 | 6 | secular states |
| 100 | 4 | 7 | muslim communities |
| 100 | 4 | 8 | christian neighbors |
| 100 | 4 | 9 | birthplace |
| 100 | 4 | 10 | Islam |
| 100 | 4 | 11 | turn |
| 100 | 4 | 12 | religion |

Fig.4 Output of noun phrase extractor for a sample input

4.3 Features, Weighting and Normalization

After identifying the document phrases, a document is reduced to a collection of noun phrases. Since, in our work, we focus on the keyphrase extraction task from scientific articles which are generally very long in size (6 to more than 20 pages), the collection of noun phrases identified in an article may be huge in number. Among theses huge collection, a small number of phrases (5 to 15 phrases) may be selected as the keyphrases. Whether a candidate phrase is a keyphrase or not can be decided by a classifier based on a set of features characterizing a phrase.

Discovering good features for a classification task is very much an art. The different features characterizing candidate noun phrases, feature weighting and normalization methods are discussed below.

**Phrase frequency, phrase links to other phrases and Inverse Document Frequency**

If a noun phrase is occurring more frequently in a document, the phrase is assumed to more important in the document. Number of times a phrase occurs independently in a document with its entirety has been considered as the phrase frequency (PF). A noun phrase may appear in a text either independently or as a part of other noun phrases. These two types of appearances of noun phrases should be distinguished. If a noun phrase P1 appears in full as a part of another noun phrase P2 (that is, P1 is contained in P2), it is considered that P1 has a link to P2. Number of times a noun phrase (NP) has links to other phrases is counted and considered as the phrase link count (PLC). Two features, phrase frequency (PF) and phrase link count (PLC) are combined to have a single feature value using the following measure:

$$F_{freq} = \sqrt{(1/2)*PF*PF + PLC}$$

In the above formula, frequency of a noun phrase (PF) is squared only to give it more importance than the phrase link count (PLC). The value 1/2 has been used to moderate the value. We explain below about this formula with an example:

Assume a phrase P1 whose PF value is 10, PLC value is 20 and PF+PLC = 30. For another phrase P2 whose PF value is 20, PLC value is 10 and PF+PLC =30. So, for these two cases, simple addition of PF and PLC do not make any difference in assigning weights to the noun phrases although the independent occurrence of noun phrase P2 is more than that of the noun phrase P1. But the independent existence of a phrase should get higher importance while deciding whether a phrase is keyphrase worthy or not. In a more general case, consider that a single word noun phrase NP1 occurs only once in independent existence and occurs (n+1) times as a part of other noun phrases and NP2 is another phrase, which occurs n times independently and occurs only once as a part of other phrases. In this situation, simple addition of PF and PLC will favor the first phrase, but our formula will give higher score to the second phrase because it occurs more independently than the first one.

Inverse document frequency (IDF) is a useful measure to determine the commonness of a term in a corpus. IDF value is computed using the formula: log(N/df), where N= total number of documents in a corpus and df (document frequency) means the number of documents in which a term occurs. A term with a lower df value means the term is less frequent in the corpus and hence idf value becomes higher. So, if idf value of a term is higher, the term is relatively rare in the corpus. In this way, idf value is a measure for determining the rarity of a term in a corpus. Traditionally, TF (term frequency) value of a term is multiplied by IDF to compute the importance of a term, where TF indicates frequency of a term in a document. TF*IDF measure favors a relatively rare term which is more frequent in a document. We combine $F_{freq}$ and IDF in the following way to have a variant of Edmundsonian thematic feature [24]:

$$F_{thematic} = F_{freq} * IDF$$

The value of this feature is normalized by dividing the value by the maximum $F_{thematic}$ score in a collection of $F_{thematic}$ scores obtained by the phrases corresponding to a document.

**Phrase Position**

If a phrase occurs in the title or abstract of a document, it should be given more score. So, we consider the position of the first occurrence of a phrase in a document as a feature. Unlike the previous approaches [14] [16] that assume the position of a phrase as a binary feature, in our work, the score of a phrase that occurs first in the sentence i is computed using the following formula:

$$F_{pos} = \frac{1}{\sqrt{i}} \text{, if i} <= n$$

, where n is the position of the last sentence in the abstract of a document. For i > n, $F_{pos}$ is set to 0.

*Phrase Length and Word Length*

These two features can be considered as the structural features of a phrase. Phrase length becomes an important feature in keyphrase extraction task because the length of keyphrases usually varies from 1 word to 3 words. We find that keyphrase consisting of 4 or more words are relatively rare in our corpus.

Length of the words in a phrase can be considered as a feature. According to Zipf's Law [21], shorter words occur more frequently than the larger ones. For example, articles occur more frequently in a text. So, the word length can be an indication for the rarity of a word. We consider the length of the longest word in a phrase as a feature.

If the length of a phrase is PL and the length of the longest word in the phrase is WL, these two feature values are combined to have a single feature value using the following formula:

$$F_{PL*WL} = \sqrt{\log(1+PL)*\log(1+WL)}$$

The value of this feature is normalized by dividing the value by the maximum value of the feature in the collection of phrases corresponding to a document.

4.4 Keyphrase Extraction Using Multilayer Perceptron Neural Network

Training a Multilayer Perceptron (MLP) Neural Network for keyphrase extraction requires document noun phrases to be represented as the feature vectors. For this purpose, we write a computer program for automatically extracting values for the features characterizing the noun phrases in the documents. Author assigned keyphrases are removed from each original document and stored in the different files with a document identification number. For each noun phrase NP in each document d in our dataset, we extract the values of the features of the NP from d using the measures discussed in subsection 4.3. If the noun phrase NP is found in the list of author assigned keyphrases associated with the document d, we label the noun phrase as a "Positive" example and if it is not found we label the phrase as a "negative" example. Thus the feature vector for each noun phrase looks like {<$a_1$ $a_2$ $a_3$ ….. $a_n$>, <label>} which becomes a training instance (example) for a Multilayer Perceptron Neural Network, where $a_1$, $a_2$ . . .$a_n$, indicate feature values for a noun phrase. A training set consisting of a set of instances of the above form is built up by running a computer program on a set of documents selected from our corpus.

After preparation of the training dataset, a Multilayer Perceptron Neural Network is trained on the training set to classify the noun phrases as one of two categories:

"Positive" or "Negative". Positive category indicates that a noun phrase is a keyphrase and the negative category indicates that it is not a keyphrase.

```
Input:

   A file containing the noun phrases of a test document
with their classifications (positive or negative) and the
probability estimates of the classes to which the phrases
belong.

   Begin:

   i. Select the noun phrases, which have been classified as
positive by the classifier and reorder these selected noun
phrases in decreasing order of their probability estimates of
being in class 1 (positive). Save the selected phrases in to an
output file and delete them from the input file.

   ii. For the rest of the noun phrases in the input file,
which are classified by the classifier as "Negative", we
order the phrases in increasing order of their probability
estimates of being in the class 0 (negative). In effect, the
phrase for which the probability estimate of being in class 0
is minimum comes at the top. Append the ordered phrases to
the output file.

   iii. Save the output file
end
```

Fig.5 Noun Phrase Ranking Based on Classifier's Decisions

For our experiment, we use Weka (www.cs.waikato.ac.nz/ml/weka) machine learning tools. We use Weka's Simple CLI utility, which provides a simple command-line interface that allows direct execution of WEKA commands.

The training data is stored in a .ARFF format which is an important requirement for WEKA.

The multilayer perceptron is included under the panel Classifier/ functions of WEKA workbench. The description of how to use MLP in keyphrase extraction has been discussed in the section 3. For our work, the classifier MLP of the WEKA suite has been trained with the following values of its parameters:

Number of layers: 3 (one input layer, one hidden layer and one output layer).
Number of hidden nodes: (number of attributes + number of classes)/2

| | |
|---|---|
| Learning rate: | 0.3 |
| Momentum: | 0.2 |
| Training iteration: | 500 |
| Validation threshold: | 20 |

WEKA uses backpropagation algorithm for training the multilayer perceptron neural network.

The trained neural network is applied on a test document whose noun phrases are also represented in the form of feature vectors using the similar method applied on the training documents. During testing, we use –p option (soft threshold option). With this option, we can generate a probability estimate for the class of each vector. This is required when the number of noun phrases classified as positive by the classifier is less than the desired number of the keyphrases. It is possible to save the output in a file using indirection sign (>) and a file name. We save the output produced by the classifier for each test document in a separate file. Then we rank the phrases using the algorithm shown in figure 5 for keyphrase extraction.

After ranking the noun phrases, K- top ranked noun phrases are selected as keyphrases for each input test document.

## 5. Evaluation and Experimental Results

There are two usual practices for evaluating the effectiveness of a keyphrase extraction system. One method is to use human judgment, asking human experts to give scores to the keyphrases generated by a system. Another method, less costly, is to measure how well the system-generated keyphrases match the author-assigned keyphrases. It is a common practice to use the second approach in evaluating a keyphrase extraction system [7][8] [11][19]. We also prefer the second approach to evaluate our keyphrase extraction system by computing its precision and recall using the author-provided keyphrases for the documents in our corpus. For our experiments, precision is defined as the proportion of the extracted keyphrases that match the keyphrases assigned by a document's author(s). Recall is defined as the proportion of the keyphrases assigned by a document's author(s) that are extracted by the keyphrase extraction system.

5.1 Experimental Dataset

The data collection used for our experiments consists of 150 full journal articles whose size ranges from 6 pages to 30 pages. Full journal articles are downloaded from the websites of the journals in three domains: Economics, Legal (Law) and Medical.

Articles on Economics are collected from the various issues of the journals such as Journal of Economics (Springer), Journal of Public Economics (Elsevier), Economics Letters, Journal of Policy Modeling. All these articles are available in PDF format.

Articles on Law and legal cases have been downloaded from the various issues of the law journals such as Computer Law and Security Review (Elsevier), International Review of Law and Economics (Elsevier), European Journal of Law and Economics (Springer), Computer Law and Security Report (Elsevier), AGORA International Journal of Juridical Sciences(Open access).

Medical articles are downloaded from the various issues of the medical journals such as Indian Journal of Medicine, Indian Journal of Pediatrics, Journal of Psychology and Counseling, African journal of Traditional, Complementary and Alternative Medicines, Indian Journal of Surgery, Journal of General Internal Medicine, journal of General Internal Medicine, The American Journal of Medicine, International Journal of Cardiology, Journal of Anxiety Disorders. Number of articles under each category used in our experiments is shown in the table 1.

Table 1: Source documents used in our experiments

| Source Document Type | Number of Documents |
| --- | --- |
| Economics | 60 |
| Law | 40 |
| Medical | 50 |

For the system evaluation, the set of journal articles are divided into multiple folds where each fold consists of one training set of 100 documents and a test set of 50 documents. The training set and the test set are independent from each other. The set of author assigned keyphrases available with the articles are manually removed before candidate terms are extracted. For all experiments discussed in this paper, the same splits of our dataset in to a training set and a test set are used. Some useful statistics about our corpus are given below.

Total number of noun phrases in our corpus is 144978. The average number of author-provided keyphrases for all the documents in our corpus is 4.90.

The average number of keyphrases that appears in all the source documents in our corpus is 4.34. Here it is interesting to note that all the author assigned keyphrases for a document may not occur in the document itself.

The average number of keyphrases that appear in the list of candidate phrases extracted from all the documents in our corpus is 3.50. These statistics interestingly show that some keyphrase worthy phrases may be missed at the stage of the candidate phrase extraction. The main problems related to designing a robust candidate phrase extraction algorithm are: (1) an irregular structure of a keyphrase, that is, it may contain only a single word or a multiword noun phrase or multiple multiword noun phrases connected by prepositions (an example of a keyphrase containing multiple multiword noun phrases is: "The National Council for Combating Discrimination"), (2) the ill-formatted input texts which are generated by a pdf-to-text converter from the scientific articles usually available in pdf format.

5.2 Experiments

We conducted two experiments to judge the effectiveness of the proposed keyphrase extraction method.

**Experiment 1**

In this experiment, we develop a neural network based keyphrase system as we discuss in this paper. All the features discussed in the subsection 4.3 are incorporated in this system.

**Experiment 2**

This is to compare the proposed system to an existing system. Kea [17] is now a publicly available keyphrase extraction system. Kea uses a limited number of features such as positional information and TF*IDF feature for keyphrase extraction. The keyphrase extraction system, Kea uses the Naïve Bayesian learning algorithm for keyphrase extraction.

We download the version 5.0 of Kea[1] and install it on our machine. A separate model is built for each fold which contains 100 training documents and 50 test documents. Kea builds a model from each training dataset using Naïve Bayes and uses this pre-built model to extract keyphrases from the test documents.

5.3 Results

To measure the overall performance of the proposed neural network based keyphrase extraction system and the publicly available keyphrase extraction system, Kea, our experimental dataset consisting of 150 documents are divided into 3 folds for 3-fold cross validation where each fold contains two independent sets: a training set of 100 documents and a test set of 50 documents. A separate model is built for each fold to collect 3 test results, which are averaged to obtain the final results for a system. The number of keyphrases to be extracted (value for K) is set to 5, 10 and 15 for each of keyphrase extraction systems discussed in this paper.

---

[1] http://www.nzdl.org/Kea/

Table 2 shows the author assigned keyphrases for the journal article number 12 in our corpus. Table 3 and table 4 show respectively the top 5 keyphrases extracted by the MLP based system and Kea when the journal article number 12 in our corpus is presented as a test document to these systems.

Table 2: Author assigned keyphrases for the journal article number 12 in our test corpus

| Dno | AuthorKey |
|---|---|
| 12 | adult immunization |
| 12 | barriers |
| 12 | consumer |
| 12 | provider survey |

Table 3: Top 5 keyphrases extracted by the proposed MLP based keyphrases extractor

| Dno | NP |
|---|---|
| 12 | immunization |
| 12 | adult immunization |
| 12 | healthcare providers |
| 12 | consumers |
| 12 | barriers |

Table 4: Top 5 keyphrases extracted by Kea

| Dno | NP |
|---|---|
| 12 | adult |
| 12 | immunization |
| 12 | vaccine |
| 12 | healthcare |
| 12 | barriers |

Table 2 and table 3 show that out of 5 keyphrases extracted by the MLP based approach, 3 keyphrases match with the author assigned keyphrases. The overall performance of the proposed MLP based Keyphrases extractor has been shown in the table 5. Table 2 and table 4 show that out of 5 keyphrases extracted by Kea, only one matches with the author assigned keyphrases. The overall performance of Kea has been compared with the proposed MLP based keyphrase extraction system in table 5.

Table 5: Comparisons of the performances of the proposed MLP based keyphrase Extraction System and Kea

| Number of keyphrases | Average Precision | | Average Recall | |
|---|---|---|---|---|
| | MLP | Kea | MLP | Kea |
| 5 | 0.34 | 0.28 | 0.35 | 0.29 |
| 10 | 0.22 | 0.19 | 0.46 | 0.40 |
| 15 | 0.17 | 0.15 | 0.51 | 0.48 |

Table 5 shows the comparisons of the performances of the proposed MLP based keyphrase extraction system and Kea.

From table 5, we can clearly conclude that the proposed keyphrase extraction system outperforms Kea for all three cases shown in three different rows of the table.

To interpret the results shown in the table 5, we like to analyze the upper bounds of precision and recall of a keyphrase extraction system on our dataset. Our analysis on upper bounds of precision and recall of a keyphrase extraction system on our dataset can be presented in two ways: (1) some author-provided keyphrases might not occur in the document they were assigned to. According to our corpus, about 88% of author-provided keyphrases appear somewhere in the source documents of our corpus. After extracting candidate phrases using our candidate phrase extraction algorithm, we find that only 72% of author provided keyphrases appear somewhere in the list of candidate phrases extracted from all the source documents. So, keeping our candidate phrase extraction algorithm fixed if a system is designed with the best possible features or a system is allowed to extract all the phrases in each document as the keyphrases, the highest possible average recall for a system can be 0.72. In our experiments, the average number of author-provided keyphrases for all the documents is only 4.90, so the precision would not be high even when the number of extracted keyphrases is large. For example, when the number of keyphrases to be extracted for each document is set to 10, the highest possible average precision is around 0.3528 (4.90 * 0.72/10 = 0.3528), (2) assume that the candidate phrase extraction procedure is perfect, that is, it is capable of representing all the source documents in to a collection of candidate phrases in such way that all author provided keyphrases appearing in the source documents also appear in the list of candidate phrases. If it is the case, 88% of the author provided keyphrases appear somewhere in the list of candidate phrases because, on an average, 88% of the author provided keyphrases appear somewhere in the source documents of our corpus. In this case, if a system is allowed to extract all the phrases in each document as the keyphrases, the highest possible average recall for a system can be 0.88 and when the number of keyphrases to be extracted for each document is set to 10, the highest possible average precision is around 0.4312(4.90 * 0.88/10 =0.4312).

## 6. Conclusions

This paper presents a novel keyphrase extraction approach using neural networks. For predicting whether a phrase is a keyphrase or not, we use the estimated class probabilities as the confidence scores which are used in re-ranking the phrases belonging to a class: positive or negative. To identify the keyphrases, we use five features such as TF*IDF, position of a phrase's first appearance, phrase length, word length in a phrase and the links of a phrase to other phrases. The proposed system performs better than a publicly available keyphrase extraction system called Kea. As a future work, we have planned to improve the proposed system by (1) improving the candidate phrase extraction module of the system and (2) incorporating new features such as structural features, lexical features.